\documentclass[sigconf]{acmart}
\AtBeginDocument{%
  }

\copyrightyear{2025}
\acmYear{2025}
\setcopyright{cc}
\setcctype{by}
\acmConference[L@S '25]{Proceedings of the Twelfth ACM Conference on Learning @ Scale}{July 21--23, 2025}{Palermo, Italy}
\acmBooktitle{Proceedings of the Twelfth ACM Conference on Learning @ Scale (L@S '25), July 21--23, 2025, Palermo, Italy}
\acmDOI{10.1145/3698205.3729542}
\acmISBN{979-8-4007-1291-3/2025/07}
\begin{document}

\title{How Adding Metacognitive Requirements in Support of AI Feedback in Practice Exams Transforms Student Learning Behaviors}

\author{Mak Ahmad}

\orcid{0000-0001-8697-2035}
\affiliation{
  \institution{University of California, Davis}
  \city{Davis}
  \state{CA}
  \country{USA}
  \postcode{}
}
\email{shahmad@ucdavis.edu}

\author{Prerna Ravi}
\orcid{0000-0002-4289-5610}
\affiliation{%
 \institution{Massachusetts Institute of Technology}
 \city{Cambridge}
 \state{MA}
 \country{USA}}
\email{prernar@mit.edu}

\author{David Karger}
\orcid{0000-0002-0024-5847}
\affiliation{%
 \institution{Massachusetts Institute of Technology}
 \city{Cambridge}
 \state{MA}
 \country{USA}}
\email{karger@mit.edu}

\author{Marc Facciotti}
\orcid{0000-0003-4453-3274}
\affiliation{%
  \institution{University of California, Davis}
  \city{Davis}
  \state{CA}
  \country{USA}}
\email{mtfacciotti@ucdavis.edu}

\renewcommand{\shortauthors}{Mak Ahmad, Prerna Ravi, David Karger, and Marc Facciotti}

\begin{abstract}
Providing personalized, detailed feedback at scale in large undergraduate STEM courses remains a persistent challenge. We present an empirically evaluated practice exam system that integrates AI generated feedback with targeted textbook references, deployed
in a large introductory biology course.   Our system specifically aims to encourage \emph{metacognitive behavior} by asking students to \emph{explain their answers} and \emph{declare their confidence}.
It uses OpenAI’s
GPT-4o to generate personalized feedback based on this information, while directing them to
relevant textbook sections. Through detailed interaction logs from consenting participants across three midterms (541, 342, and 413 students respectively), totaling 28,313 question-student interactions across 146 learning objectives, along with 279 post-exam surveys and 23 semi-structured interviews, we examined the system's impact on learning outcomes and student engagement. Analysis showed that across all midterms, the different feedback types showed no statistically significant differences in performance, though there were some trends suggesting potential benefits worth further investigation. The system’s most substantial impact emerged
through its required confidence ratings and explanations, which
students reported transferring to their actual exam strategies. Approximately 40\% of students engaged with textbook references
when prompted by feedback---significantly higher than traditional
reading compliance rates. Survey data revealed high student satisfaction (M=4.1/5), with 82.1\% reporting increased confidence on
midterm topics they had practiced, and 73.4\% indicating they could
recall and apply specific concepts from practice sessions. Our findings demonstrate how thoughtfully designed AI-enhanced systems
can scale formative assessment while promoting sustainable study
practices and self-regulated learning behaviors, suggesting that embedding structured reflection requirements may be more impactful
than sophisticated feedback mechanisms.
\end{abstract}






\maketitle

\section{Introduction}

It is a truism among faculty that students do not read the textbook \cite{clump2004extent, french2015textbook}. In contrast, our experience is that students are strongly motivated to tackle practice exams in the run-up to actual exams \cite{carpenter2012testing, roediger2011critical}. This paper describes an experimental deployment, in a class of 1002 students, of a system we developed that aimed to turn practice exams into a teaching tool. While the AI features meant to be the system's centerpiece did not drive improvements in student exam performance, two other deliberate design choices aimed at shaping student engagement with the AI feedback---requiring students to 1) rate their confidence and 2) explain their reasoning while attempting every question---were unexpectedly impactful.
These features had unanticipated positive effects on students' metacognition, internal motivation, and long-term exam preparation strategies. 

To make multiple-choice practice exams more interactive we developed a tool that could give students \emph{formative feedback}~\cite{hattie2007power, kulhavy1985feedback} on their answer choices, rather than simple right and wrong responses. We used a Large Language Model (LLM) to either (i) guide students to an appropriate place in the textbook to acquire the knowledge they needed to answer correctly or (ii) \emph{generate} textual guidance about the material the student was getting wrong. To make this feedback more targeted, our design required that each student \emph{explain} their answer when they submitted it; this explanation could help the LLM understand what the student did not understand, in order to steer them to the right textbook section or to generate suitable corrective guidance. Based on analysis of prior educational feedback literature, we also surmised that it would be beneficial to record students' \emph{confidence} in their answer, and use that to further customize the feedback \cite{kulhavy1985feedback,hattie2007power}.

We ran a randomized controlled trial in a 1002-student general biology class, comparing these two feedback approaches (guide-to-textbook and AI-generated feedback) to a control that only told students whether their answers were correct. Contrary to our expectations based on prior work on elaborated feedback, we found no significant difference between treatment conditions. Instead, the most substantial impact came from the metacognitive requirements we had initially included as supporting elements for the feedback system. This finding suggests that structured reflection may be more valuable for student learning than the specific content of feedback provided.

While the benefits of practice testing are well-documented \cite{carpenter2012testing, roediger2011critical}, providing effective, personalized feedback to large student populations remains a significant challenge \cite{butler2007effect,wisniewski2020power}. The emergence of sophisticated AI systems capable of generating contextual feedback offers promising new approaches to this challenge \cite{tang2021does,ward2024analyzing}, especially when considered within the framework of self-regulated learning and student success \cite{hattie2007power,wisniewski2020power}, though careful integration is needed to balance AI assistance with student agency \cite{ahmad1}.

Traditional practice exam approaches face several key limitations. Static feedback and delayed grading often fail to provide students with the timely guidance they need \cite{kulik1988timing,butler2007effect}, while faculty time constraints make it difficult to offer detailed, personalized feedback to hundreds of students \cite{smolansky2023educator}. Furthermore, students frequently focus on their numerical grades rather than developing deeper understanding of the course material \cite{wisniewski2020power,vafeas2013attitudes}. This tendency is exacerbated by the perceived disconnect between exam performance and specific textbook content, making it challenging for students to effectively address their knowledge gaps \cite{klymkowsky2024end,chipperfield2022embedding}.

Because students are motivated by practice exams, we aimed to turn them into better teaching tools by providing customized feedback on student responses.  The theory of effective feedback systems draws from several key education research areas. Self-regulated learning theory emphasizes the importance of helping students develop metacognitive strategies and self-assessment skills \cite{hattie2007power, wisniewski2020power}. Student confidence plays a crucial role in both learning and assessment, influencing how feedback is received and integrated~\cite{hattie2007power, kulhavy1985feedback}.  This work suggests that effective feedback systems should not only provide accurate information, but also support students in developing better study strategies and engaging more meaningfully with course materials.

Despite growing interest in AI-enhanced educational tools, significant gaps remain in our understanding of their effectiveness. Limited research exists on AI-generated feedback in authentic learning contexts, particularly regarding how different types of feedback compare in supporting student learning. While research has established a connection between student confidence and the effectiveness of feedback \cite{hattie2007power, kulhavy1989feedback}, its application in the context of AI-powered feedback remains unexplored. While research has established ways to improve alignment between instruction and assessment \cite{chipperfield2022embedding,biggs2011train}, the impact of directly connecting feedback to specific textbook content remains unexplored, particularly in the context of AI-enhanced learning systems at scale.

To address these gaps, this study investigates two areas:

\textbf{RQ1:} Does supplementation of traditional midterm exam study resources with an AI-powered practice exam system with textbook linking impact student exam performance and related study behaviors?

\textbf{RQ2:} How do design elements of the system accompanying AI feedback influence student attitudes, engagement, and self-regulated learning behaviors?

\textbf{This study introduces a novel approach to practice exam feedback by combining AI-generated explanations with direct textbook references, implemented in a large enrollment undergraduate biology course. The use of required confidence ratings and student explanations as personalized inputs for the LLM also introduces a new dimension to the AI feedback literature.}
By examining both learning outcomes and student engagement patterns,  we contribute to the understanding of how technology can enhance educational feedback at scale. 

\section{Related Work}

\subsection{Feedback in Educational Settings}

Prior work has established foundational insights into how \textit{timing} and \textit{type} of feedback could influence the quality of learning.

%  FEEDBACK TYPE PARA
Studies have shown that merely indicating correctness provides insufficient scaffolding for conceptual growth, whereas elaborative feedback explaining \textit{why} an answer was correct or incorrect fostered better retention and application \cite{kulhavy1985feedback}. 
Researchers have devised models to assess effective feedback using three major questions asked by teachers and/or students:  1) \textit{Where am I going?} (goal alignment), 2) \textit{How am I going?} (progress evaluation), and 3) \textit{Where to next?} (future guidance) \cite{hattie2007power}. This framework emphasizes moving beyond binary correctness indicators to address cognitive processes and self-regulation, aligning feedback with learners’ developmental trajectories. This distinction has been corroborated by longitudinal cognitive diagnostic assessments, which found that cognitive diagnostic feedback tailored to learners’ mastery of specific attributes significantly outperformed traditional correct-incorrect response feedback in challenging domains \cite{tang2021does}.  Similarly, studies have shown that high-information feedback (combining task-level correctness with self-regulation from monitoring attention, emotions, or motivation during learning) yielded better effects than simple forms of reinforcement or punishment at the task level \cite{wisniewski2020power, hattie2007power, lysakowski1981classroom}. Such findings underscore the importance of granular, attribute-level feedback in addressing knowledge gaps. 

% FEEDBACK TIMING PARA
Timing further modulates feedback's utility \cite{hattie2007power}. Immediate feedback seems more effective for learning procedural skills and correcting errors in practice, while delayed feedback appears to better support long-term retention of conceptual knowledge \cite{butler2007effect,kulik1988timing, tang2021does}. Finally, learner confidence and self-assessment profoundly influence feedback efficacy, as overestimations or underestimations of competence can distort receptivity. Researchers found that students who were confident in their answers benefited most from simple feedback about correctness, while students with low confidence needed more detailed guidance explaining underlying concepts and connecting them to foundational knowledge they might be missing \cite{hattie2007power, kulhavy1989feedback}. Furthermore, research has shown that self-regulated learners—those adept at monitoring and adjusting their strategies—derive more benefit from feedback than peers reliant on external guidance, emphasizing the interplay between feedback design and metacognitive skills \cite{butler1995feedback}. This approach leverages metacognitive awareness, enabling learners to reconcile discrepancies between perceived and actual understanding.

Our study builds upon these foundational insights by integrating automated feedback with confidence-aligned strategies and textbook references, aiming to provide personalized, high-information feedback at scale while promoting self-regulated learning behaviors in a large undergraduate biology course. 

\subsection{Textbook Reading Practices in Higher Education}

%  why students dont like reading assignments
 While academic reading fosters disciplinary discourse and improves writing skills \cite{howard2018academic, sengupta2002developing, lockhart2016critical}, studies across disciplines reveal low compliance with reading assignments, with as few as 27\% of students completing readings before class \cite{clump2004extent} and 72\% admitting to rarely or never reading on schedule \cite{st2018exploring, connor2000assessing}. Reasons for this include competing demands on time, lack of motivation, and misalignment between student and faculty expectations \cite{starcher2011encouraging, nolen1996study, rothkopf1988perspectives}. While motivated students with a high need for cognition or mastery-oriented goals may engage with textbooks more deeply \cite{derryberry2008relationships, starcher2011encouraging}, others perceive reading as effortful and unnecessary if lectures or other resources provide sufficient information \cite{sappington2002two, murden1997role, st2018exploring,vafeas2013attitudes}. Furthermore, students struggle when faculty do not explicitly integrate assigned readings into classroom discussions or assessments, reinforcing the perception that reading is optional \cite{brost2006student, starcher2011encouraging, maher2010m,french2015textbook, hoeft2012university}.

 %  solutions devised by researchers for increasing reading engagement
Scholars have devised ways to improve student engagement with reading assignments. Prior work has found that introducing just-in-time quizzes during reading or in-class quizzes following a chapter's reading assignment 
can improve reading habits \cite{, howard2004just, st2018exploring,hatteberg2013increasing, heiner2014preparing, johnson2009effect}. Beyond quizzes, question-based approaches foster self-assessment; research has also found that asking students to pose a question after reading the textbook enhances comprehension and leads to better exam performance \cite{van2006study, marbach2000can}.
Critical scoped reading has also been emphasized through structured prompts \cite{tomasek2009critical}. Different reading formats have also been explored. 
While some studies suggested readers (a thematic compilation of  excerpts from different authors grouped to explain a topic), over traditional textbook formats for increasing engagement and critical thinking \cite{howard2004just}, others found textbook format choice had little impact \cite{durwin2008does}. Other approaches, such as experiential learning models \cite{stokes2007using} and self-monitoring strategies \cite{chang2010effects}, also aim to enhance reading compliance and academic performance.

% gaps between readings and assessments 
Current practices also reveal fundamental gaps between course materials and assessments. Recent work highlights that assessment and instruction are often conceived as separate in both time and purpose: without explicit links between both, students struggle to recognize how instructional materials—including readings—contribute to building the skills needed for success on exams \cite{klymkowsky2024end}. The constructive alignment framework \cite{biggs2011train} emphasizes the need for assessments that are directly tied to instructional content, ensuring coherence between teaching methods, evaluation, and learning objectives. Embedding instructional resources directly into assessments may also provide students with immediate access to relevant materials at critical learning moments \cite{chipperfield2022embedding}, addressing the challenge of expecting students to independently seek out these resources. Strengthening the link between course materials, assessments, and feedback can support deeper engagement and knowledge retention.

Our study addresses these gaps by embedding links to relevant textbook sections within practice assessments, alongside AI-generated personalized feedback. We frame these references as optional, goal-oriented learning supports—designed not as required reading, but as resources students can actively use to deepen their understanding of course objectives and see clear, actionable connections to the exam questions they find most relevant.

\subsection{AI Systems in Educational Assessment}

% GenAI + personalized learning
Researchers found that generative AI offers the potential to deliver immediate and diverse feedback, along with opportunities for self-assessment, across a wide variety of learning contexts. This can encourage students to study independently and develop strong self-regulation,  including goal setting, self-monitoring, self-assessment, and adaptive learning strategies \cite{xia2024scoping}. For example, recent papers argue that AI-powered assessment tools can analyze student explanations, identifying concepts and scientific principles that may be missing or misapplied, while also making suggestions for how instructors can use these data to better guide student thinking \cite{klymkowsky2024end}.  LLM-powered reflection prompts have been shown to significantly improve student learning outcomes at scale \cite{kumar2024supporting}.
Additionally, researchers reported that AI-based tools enhance real-time formative feedback by supporting personalized learning, adaptive test adjustments, and real-time classroom analysis, with students expressing strong support for these capabilities \cite{ward2024analyzing}. 

% GenAI pitfalls
Despite these potential benefits, AI-powered feedback presents several challenges. AI systems may assess student responses using criteria different than course instructors', potentially leading to misaligned feedback without sufficient course-specific context \cite{nikolic2023chatgpt, xia2024scoping}. Students who use LLMs as personal tutors by conversing about the topic and asking for explanations benefit from usage, whereas learning is impaired for students who excessively rely on LLMs to solve practice exercises for them \cite{lehmann2024ai}. This suggests that AI-powered feedback is most effective when used as a supplement to student engagement rather than as a replacement for active learning. Excessive reliance on AI feedback can discourage students from developing important critical thinking skills through independent problem-solving \cite{jensen2024generative}.  A recent study also showed AI feedback deployed at scale that addresses several effective feedback components at the task and process level but still misses self-reflection feedback elements needed to motivate students \cite{dai2024assessing}. 

% Gaps in current AI feedback studies
Existing studies on AI-generated feedback in education also face limitations due to small sample sizes \cite{geerling2023chatgpt, stutz2023ch}, indicating the challenge of assessing the generalizability of AI-based interventions in different educational contexts. Several challenges complicate large-scale deployment of AI-powered assessment tools: instructors need time to learn and integrate new assessment types into their courses, technical infrastructure must be reliable and accessible to all students, and the systems must maintain consistent performance throughout the term  \cite{smolansky2023educator}. The need for instructors to verify the accuracy and pedagogical appropriateness of AI-generated feedback increases resource demands when implementing personalized feedback at scale \cite{nikolic2023chatgpt}. 

In this paper, we present a novel practice exam tool for midterm preparation that combines AI-generated feedback with self-regulatory learning components. We propose strategies to balance automation with pedagogical rigor, ensuring that AI-driven feedback enhances learning through structured metacognitive elements like confidence ratings and reasoning explanations.
Analyzing the effects of AI-generated feedback at scale, we build on previous controlled studies of automated feedback \cite{leite2020effects}.

\section{System Design and Implementation}

\begin{figure}[ht]
\centering
\includegraphics[width=0.8\columnwidth]{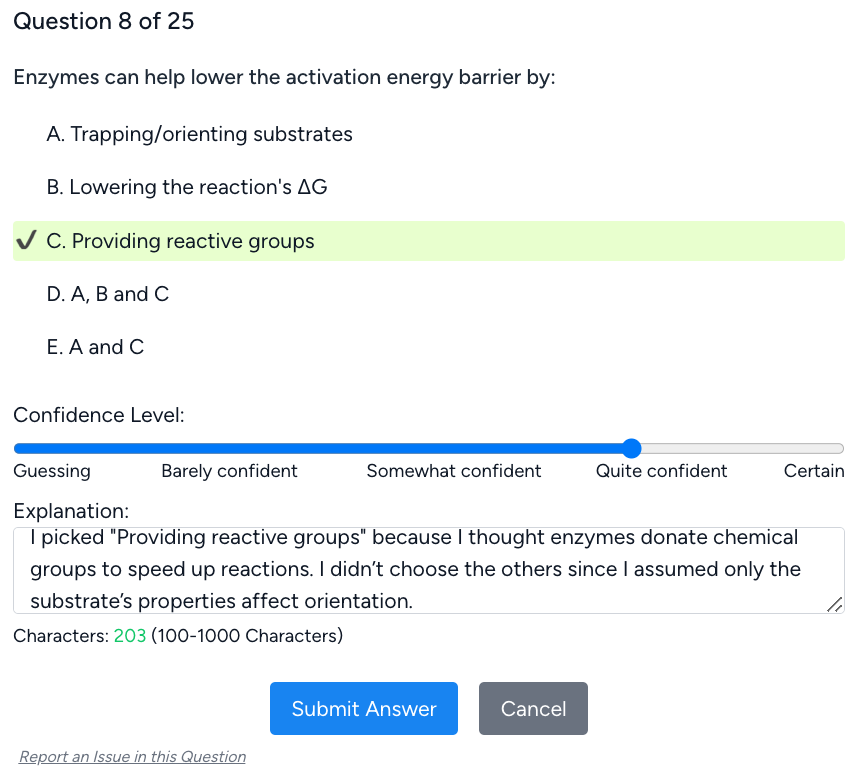}
\caption{Test mode interface where students provide their answer, confidence rating, and reasoning.}
\label{fig:test-mode}
\end{figure}

\begin{figure}[ht]
\centering
\includegraphics[width=0.8\columnwidth]{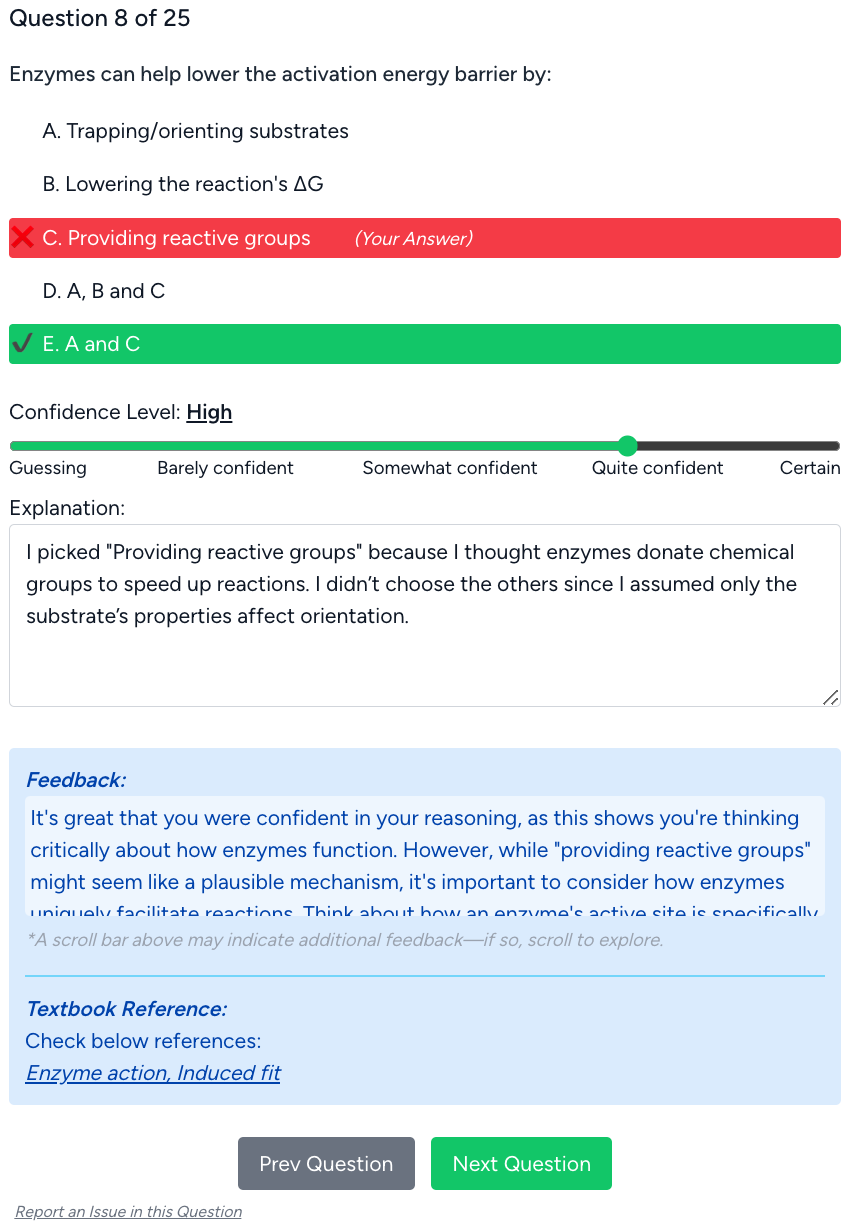}
\caption{Review mode interface showing condition-specific feedback and textbook references in a scrollable panel.}
\label{fig:review-mode}
\end{figure}

Our system aimed to provide a deeper version of the typical interactive practice exam experience for a 1002-student undergraduate biology course.  The designed workflow is as follows:
\begin{enumerate}
    \item We created a \emph{question bank} of 400 multiple-choice practice exam questions (150 image-based) across the three midterms, each labeled with relevant \emph{learning objectives}.   We then prepared a small number of practice exams each drawing 25 questions from the question bank.
    \item Given a particular practice exam, in \emph{test mode} (Figure \ref{fig:test-mode}) students answer each question.  In addition to choosing a multiple choice answer, students must indicate a \emph{confidence level} and provide an \emph{explanation} for their choice
    \item In \emph{review mode} (Figure \ref{fig:review-mode}) the system provides feedback on each answered question.   In addition to reporting right-or-wrong, the system can provide \emph{textbook links} to relevant content, and/or \emph{custom feedback} generated by an LLM from the student's choice, confidence, and explanation.
\end{enumerate}

\subsection{Question Bank Development and Learning Objectives}
We worked with the course instructor to validate learning objective mappings for approximately 400 multiple-choice questions from previous exams, including 150 image-based questions. To ensure reliable AI feedback generation, we identified relevant textbook sections and generated comprehensive summaries for each learning objective using Claude 3.5 Sonnet, which were then validated by the instructor. These summaries and textbook mappings were included in feedback prompts to constrain GPT-4o's responses to validated biological content rather than its general knowledge, significantly reducing the risk of hallucination. The entire data preparation process took over 30 hours.

\subsection{Practice Exam Implementation and Deployment}
The practice exam structure mirrored the actual midterm format while incorporating experimental elements. Each practice exam contained 25 questions, carefully selected to ensure coverage of all relevant learning objectives while maintaining the integrity of the condition assignment system. Students were required to provide both a confidence rating and written explanation for each answer before proceeding.
Motivated by the related work on effective feedback timing, we implemented a two-phase approach. In the initial attempt phase, students answered all questions without immediate feedback. Upon completing the entire exam, students entered a review mode that presented their responses alongside condition-specific feedback. This design choice was informed by research suggesting benefits of delayed feedback in certain learning contexts.

\subsection{Feedback Generation}
The AI feedback component used OpenAI's GPT-4o API. We selected GPT-4o as it is currently one of the most capable publicly available language models, with documented performance on complex reasoning tasks and educational content generation. The feedback generation pipeline begins by assembling a comprehensive context that includes the question content, correct answer, student's selected answer and explanation, relevant learning objective summaries, and the student's reported confidence level. This context feeds into a carefully engineered prompting system that constrains responses to 5-7 sentences while ensuring pedagogically appropriate tone and relevant textbook references.

The base prompt established guidelines for encouraging, constructive feedback that addressed specific elements of student reasoning without directly stating correct answers for incorrect responses. Feedback varied based on both correctness and confidence level - for instance, high-confidence incorrect answers received feedback that gently highlighted inconsistencies while guiding students to identify specific logical flaws. For low-confidence correct answers, the system focused on building confidence by connecting correct intuitions to fundamental principles. All prompts underwent multiple iterations of testing with course instructors to ensure alignment with course objectives.

To maintain reliability, the system implements robust error handling and fallback mechanisms. When API responses fail or take too long, the system gracefully degrades to simpler feedback modes (changing the experimental condition) rather than leaving students without guidance. The presentation layer implements a scrollable interface with careful attention to user experience, including visual cues encouraging students to engage with the full feedback content.

\subsection{Interaction Tracking and Analytics}
The system implements a comprehensive interaction tracking to understand how students engage with different types of feedback. Beyond basic metrics like time spent per question, we designed specific mechanisms to capture detailed interaction patterns. The feedback interface uses a deliberately constrained window height, enabling us to track whether students initiate any scrolling and whether they reach the end of the feedback content. For each question attempt, the system records temporal data between initial question load and final submission, as well as any revised answers and explanation changes during review mode. To understand varying study contexts, the system tracks device types and browsers used in both test and review modes, allowing analysis of different access patterns across these phases. All interaction data, including mouse hover patterns and textbook reference engagement, is stored in a structured format with timestamps and session identifiers. This data model enables correlation between interaction patterns and learning outcomes while controlling for variables like confidence levels and prior performance, all while maintaining reasonable storage requirements for our large student population.

\subsection{Infrastructure and Technical Stack}

\begin{figure}[ht]
\centering
\includegraphics[width=1\columnwidth]{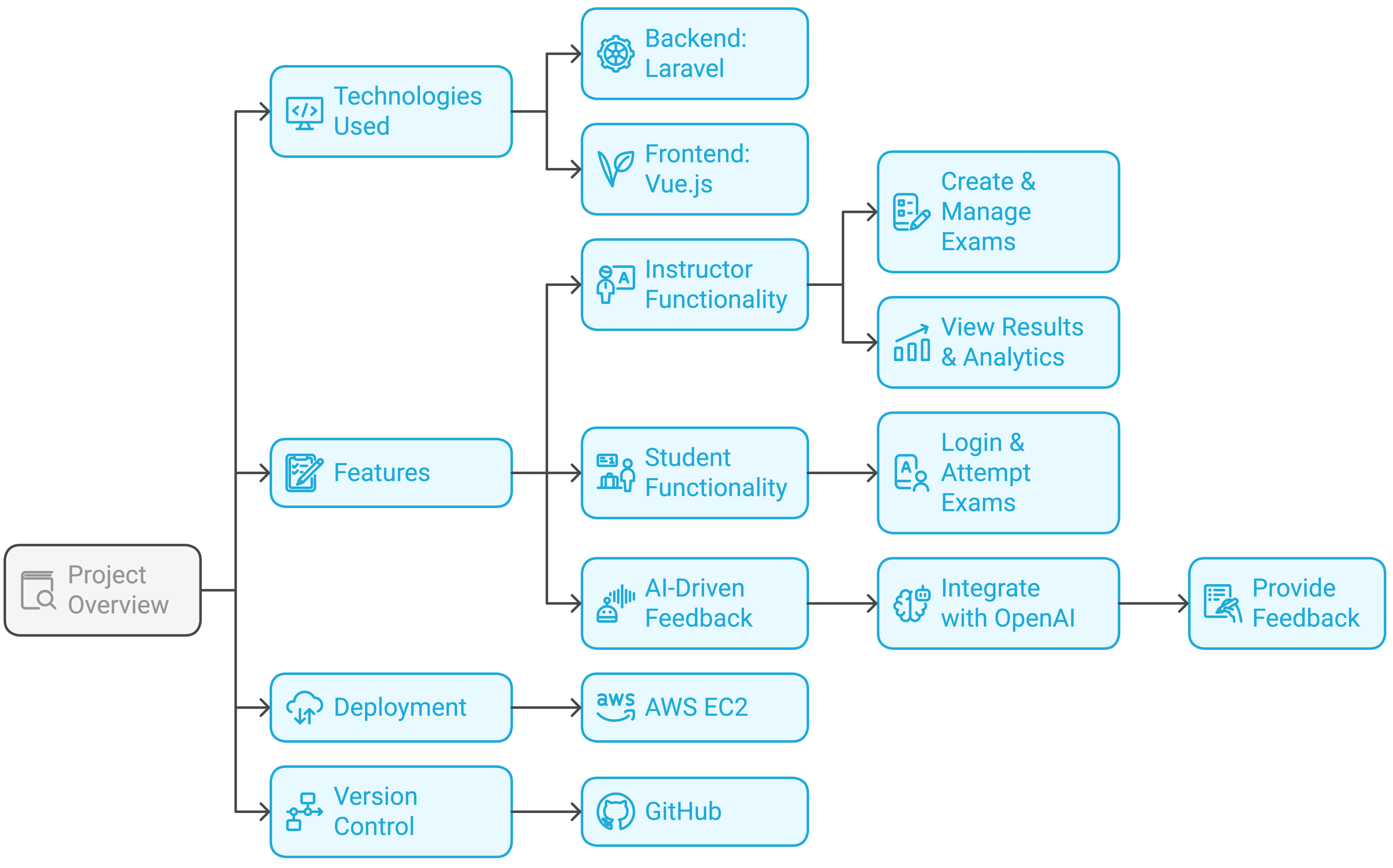}
\caption{Technologies used to build our system.}
\label{fig:system}
\end{figure}

The application (shown in figure \ref{fig:system}) was built using Laravel 11.34 and Vue.js, hosted on AWS infrastructure. The data model centered on four key entities: a question bank containing 400 multiple-choice questions with associated metadata, learning objectives mapped to textbook sections, student profiles managing authentication and experimental conditions, and comprehensive interaction logs. Question images were stored in a GitHub repository for version control and reliable delivery. The backend uses PHP 8.3.14, MySQL 8.0.40, and Nginx for web serving. The system used Google OAuth integration configured for university email addresses, ensuring secure access while minimizing friction in the student experience. This approach maintained consistent experimental conditions across sessions while restricting access to enrolled students.

\section{Experimental Methodology}

\subsection{Study Context and Participants}
This study was conducted in a first-year undergraduate general biology course offered at UC Davis.  The course has 3 midterm exams.  Use of the practice tool was assigned for credit as preparation for midterms 1 and 3 and made optional for midterm 2.

We studied 1002 students enrolled in a single offering of the course.  The practice exam tool was deployed one week before each midterm, allowing students sufficient time to engage with the platform while maintaining proximity to the actual assessment. Prior to implementation, the IRB of the hosting institution deemed the project exempt from full review (IRB 1456274-2). All participating students consented to their data being used for research purposes.

In total, 836 students (541 consenting to research) used the system for midterm 1, 760 students (342 consented) voluntarily used it for midterm 2, and 877 students (413 consented) for midterm 3, generating 28,313 question-student interactions. The high ongoing participation demonstrated sustained engagement with the tool.

\subsection{Experimental Design and Condition Assignment}

We ran a randomized controlled trial over four feedback conditions. Condition 1 only indicates whether the student's answer was right or wrong. Condition 2 augments the responses in the review phase with links to specific relevant (online) textbook sections. Condition 3 in the review phase delivers AI-generated personalized feedback using GPT-4o, factoring the student's explanation and confidence level to provide targeted guidance. Condition 4 provides both the textbook links of Condition 2 and the personalized feedback of Condition 3.  Questions were carefully distributed across conditions to ensure each student experienced all feedback types while maintaining learning objective independence.

The system maintains consistent assignment of experimental conditions. Instead of randomly assigning conditions each time a student interacts with a question, the system implements a deterministic assignment mechanism based on a hash function that considers the IDs of student, question, and practice exam. This ensures that students consistently experience the same experimental conditions for specific questions across multiple sessions, while maintaining an even distribution of conditions across students.

\subsection{Quantitative Analysis Framework}
Our analysis examined feedback condition impacts on student performance through both ANOVA and linear regression. The primary unit of analysis was student performance on individual midterm questions, mapped to practice questions through learning objectives. We employed one-way ANOVA with Tukey's HSD test ($\alpha = 0.05$) to identify differences between conditions, supplemented by a linear regression model accounting for practice exposure. The model used question-level scores as the dependent variable, with condition dummy variables and an exposure measure (number of relevant practice questions attempted per learning objective) as predictors.

To analyze how practice exam performance predicted real exam success, we mapped questions through learning objectives, with each question testing multiple objectives (average 2.3 per question). We weighted objectives equally within questions and calculated overlap weights between practice and real exam questions based on shared objectives. For example, if a practice question tested three objectives (GC.31, GC.32, GC.33) and a real exam question tested two of these (GC.31, GC.32), the overlap weight would be 0.67, representing the proportion of shared objectives.

\subsection{Survey Assessment}
A post-midterm survey (n=279, 25.8\% response rate) collected data across four categories: study habits and tool usage, feedback effectiveness, learning behaviors, and tool experience. Questions included multiple-choice, 5-point Likert scales, and open-ended responses to assess both quantitative impact and qualitative experiences with the system.

\subsection{Qualitative Interview Study}
To understand how students integrated the practice exam system into their study processes, the two lead authors conducted 23 semi-structured remote interviews with 19 participants via Zoom, each lasting 30-45 minutes. Four participants were interviewed twice-- during and after the course-- to examine whether and how their experiences with the tool influenced future study practices. We recruited these students via course announcements and incentivised participation with a \$10 gift card. Students in our interview sample had varied backgrounds in their motivations for taking the course, including fulfilling major requirements, preparing for medical school, or exploring biology as a field. Their grade expectations ranged from simply passing with a C to aiming for a high A. Additionally, they differed in their preferred exam preparation strategies, the resources they prioritized, and the timing of their practice exam usage within their study schedules. The interview protocol explored five areas: study strategies and resource usage for exam preparation, tool usage and experience when attempting the exam, feedback and learning, perceived impact on midterm performance, and feature suggestions for future integration. We investigated how students engaged with different feedback types, their experience with confidence ratings and explanation requirements, and instances where the system influenced their understanding. Follow-up interviews with the four returning participants allowed us to investigate the persistence of study behaviors across courses and the tool’s long-term influence on metacognitive study habits.

Following the interviews, the two authors conducted an inductive thematic analysis of the transcripts, allowing themes to emerge organically from the data \cite{charmaz2008grounded, corbin1990basics}. The process began with an initial review of the transcripts, during which notes were taken. A second reading facilitated the development of preliminary codes refined through discussion and iteration between the authors. After coding, related quotes were organized into overarching themes and categories. These themes and their definitions were further reviewed and refined until consensus and saturation were achieved. This qualitative analysis provided valuable context for interpreting the quantitative findings from the survey and system interaction logs.

\begin{table}
    \footnotesize
    \centering
    \begin{tabular}{|p{0.3\linewidth}|p{0.6\linewidth}|} \hline 
         \textbf{Themes}& \textbf{Definition}\\ \hline 
 \multicolumn{2}{c}{\textbf{A: Background Knowledge}}\\\hline  
         Course Importance \& Motivation& How students perceive the importance of [Anonymous course] in their academic and career trajectory, including their motivation for taking the class and desired grade outcomes.\\ \hline 
         Study Strategies \& Preparation Methods& The approaches students take to prepare for the midterm, including study materials, resources, and scheduling (including when they used the practice exam).\\ \hline 
         Identifying Knowledge Gaps \& Study Challenges& The ways students assess their understanding and the difficulties they encounter while studying.\\ \hline 
         General Role/Importance of Practice Exams& How students use practice exams (both PDF and online tool) to test their knowledge and adjust their study strategies.\\ \hline 
         PDF vs. Online Tool Preferences& The perceived advantages and disadvantages of the PDF practice test compared to the online tool.\\ \hline 
 \multicolumn{2}{c}{\textbf{B: Perceptions and Self-regulation during ATTEMPT/TEST Mode}}\\ \hline 
         Expectations \& First Impressions of the Online Tool& Students’ initial reactions to using the online practice tool and how it compared to their expectations.\\ \hline 
         Confidence Ratings Perception during ATTEMPT mode& The impact of rating confidence levels (e.g. on self-awareness) when attempting the question\\ \hline 
         Explanation Requirement Response& Details on perception towards writing down thought process and its usefulness\\ \hline 
         Confidence Ratings Use during REVIEW mode& The impact of confidence levels on error correction, feedback attention, perceived knowledge mastery, etc. when reviewing feedback\\ \hline 
 \multicolumn{2}{c}{\textbf{C: Engagement with Feedback and Learning Resources during Review Mode}}\\ \hline
         Perceived effectiveness of AI generated feedback&	Student engagement with AI-generated explanations and whether they found them useful for understanding concepts. Including anecdotes\\ \hline
         Feedback Awareness Gap&	Student understanding (or lack thereof) about feedback availability being condition-dependent; includes misconceptions about technical issues vs. experimental conditions\\ \hline
         Engagement with Textbook Links&	How and when students used the textbook links provided within the tool. Including anecdotes\\ \hline
        Engagement with External Resources&	Instances when students tried to seek additional explanations/resources outside of the tool\\ \hline 
 \multicolumn{2}{c}{\textbf{D: Perceived Effects on Midterm and Knowledge Transfer}}\\ \hline
        Perceived Midterm Performance&	The extent to which students felt that the practice exams prepared them for the actual midterm.\\ \hline
        Learning Transfer Evidence&	Specific instances where tool usage directly aided exam performance; includes concept recognition and application\\ \hline 
 \multicolumn{2}{c}{\textbf{E: Future Directions}}\\ \hline
        Usability \& Technical Challenges with the Tool&	Details on tool’s functionality and ease of use\\ \hline
        Future Study Approaches&	How using the tool influenced students’ preparation for future exams and courses.\\ \hline
        Student-Driven Suggestions for Tool Improvement&	The changes students feel would make the tool more effective.\\ \hline
    \end{tabular}
    \caption{Final set of interview categories (A-E), themes under each, and their definitions}
    \label{tab:themes}
\end{table}

\section{Results}

\subsection{Model Diagnostics}
Diagnostic tests indicated minimal autocorrelation (Durbin-Watson = 1.764) and acceptable multicollinearity (condition number = 10.8). While the Jarque-Bera test (JB = 2846.646, p < 0.001) suggested non-normality in the residuals, this doesn't invalidate our regression results given our large sample size (n = 10,820). With samples this large, the sampling distribution of regression coefficients approaches normality regardless of residual distribution, allowing for reliable statistical inference despite the normality violation.

\subsection{RQ1: Impact on Performance}
Analysis of variance across all three midterms showed no consistently significant differences between feedback conditions ($F = 2.139$, $p = 0.093$). While pairwise comparisons using Tukey's HSD test suggested some differences between conditions, these did not reach statistical significance when considering the complete dataset.

The linear regression model using Condition 1 as baseline indicated a positive trend for Condition 2 (textbook references) compared to basic feedback, equivalent to approximately 0.7\% improvement in performance, though this difference was not statistically significant ($\beta = 0.0073$, $p = 0.328$). Similarly, other conditions showed trends but no statistically significant differences. Mean scores across all midterms by condition were: 
\begin{itemize}
    \item Condition 1 (basic feedback): $M = 0.730$, $SD = 0.444$, $n = 7698$
    \item Condition 2 (textbook references): $M = 0.739$, $SD = 0.439$, $n = 6917$
    \item Condition 3 (AI feedback): $M = 0.745$, $SD = 0.436$, $n = 6723$
    \item Condition 4 (AI + textbook): $M = 0.726$, $SD = 0.446$, $n = 6975$
\end{itemize}

The most consistent finding across all midterms was that student confidence ratings strongly predicted performance ($\beta = 0.063$, $p < 0.001$), with higher confidence associated with approximately 6.3\% better performance on exam questions. This robust relationship reinforces the value of metacognitive elements in the system.

A separate analysis of lower-performing students (those scoring in the bottom 20\% on initial practice exams) showed a potentially different pattern. While the overall model approached but did not reach statistical significance ($F = 2.48$, $p = 0.059$), there was a trend suggesting that Condition 4 (AI + textbook) might provide additional benefit compared to baseline ($\beta = 0.049$, $p = 0.067$). This represents a potential performance improvement equivalent to approximately one additional correct question on a 20-question exam, and could indicate that combined feedback might be more beneficial for students who would benefit from additional learning support—a hypothesis warranting further investigation.

\subsubsection{Survey Results}
Based on 279 survey responses, students rated different feedback types on a 5-point scale. Correctness feedback received the highest average rating (\(M=3.82\), \(SD=0.94\)), followed by combined AI and textbook feedback (\(M=3.52\), \(SD=1.13\)), AI-generated explanations alone (\(M=3.44\), \(SD=1.08\)), and textbook references (\(M=3.41\), \(SD=1.12\)). A strong majority (82.1\%) of survey respondents  reported increased confidence on the midterm for topics they had practiced using the system. When asked about concept recall, 73.4\% of survey respondents indicated they could specifically remember and apply concepts from their practice sessions during the actual midterm.

\subsection{RQ2: Impact on Engagement and Learning Behaviors}

\subsubsection{Survey Results}
Following midterm 1, we conducted a post-exam survey that received 279 responses (25.8\% response rate). These survey results reflect student experiences with the system during preparation for the first midterm examination only.

Of the total 1002 enrolled students, 804 (80.2\%) used the online practice exam tool for midterm 1, with 541 consenting to have their data included in the research. Among the survey respondents, 65.9\% reported using both the online tool and a separate PDF practice exam, while 26.9\% reported using only the online tool. The PDF practice exam contained different questions than the online version, offering complementary practice opportunities.

Regarding tool satisfaction after midterm 1, students rated the system's ease of use favorably ($M=4.1/5$, $SD=0.89$). 76.3\% of survey respondents indicated they would use it again for future exams - a prediction validated by the 76\% voluntary adoption rate when the tool became optional for midterm 2. The most frequently requested improvement was enhanced AI explanations (68.4\% of respondents), particularly for complex topics and incorrect answers.

Student survey responses indicated changes in study behaviors after using the online tool for midterm 1. Among those who used the tool, 73.4\% reported adjusting their study approaches based on the feedback received. Analysis of midterm 1 interaction data showed that approximately 28\% of students clicked on provided textbook links, with similar rates between those who received only textbook references and those who received both AI feedback and textbook references. By midterm 3, this engagement with textbook references increased to approximately 39\%.

\subsubsection{Qualitative insights from interviews} For this paper's RQs, we focus on themes from categories B, C, and D (See Table \ref{tab:themes}) below.

\paragraph{\textbf{Student perceptions and self-regulation during test mode:}} 

The integration of confidence ratings and explanation requirements in the online practice tool played a significant role in shaping students’ self-assessment, engagement with the material, and overall learning experience. Their effectiveness however varied based on individual study habits, highlighting both strengths and challenges of incorporating structured reflection into an assessment tool.

% Many students were initially surprised by the requirement to both rate their confidence and provide a written explanation for their answers. They noted that they did not expect to have to articulate their reasoning in such detail, with P1 specifically highlighting the unexpected 100-character minimum before moving on. The confidence rating feature also caught some students off guard. P6 admitted that they did not initially understand its purpose but later found value in reviewing discrepancies between their confidence levels and their actual correctness. For instance, they noted that there were times when they had marked a low confidence level but ended up being correct, or conversely, felt highly confident but answered incorrectly. This realization helped them recalibrate their self-perception and accuracy in assessing their knowledge.

For many of the interviewees, the explanation requirement functioned as a tool for deeper thinking and self-reflection. P6 found that it forced them to justify their choices rather than rely on intuition, stating, \textit{``Sometimes I’d think, I’m pretty sure it’s this,’ but then I’d realize, I don’t actually have a reason to think that.'' }Similarly, others noted that the process of writing out explanations made them question their assumptions and reconsider their answers, and distinguish between educated guesses and actual understanding, a level of engagement they did not typically experience when using traditional PDF practice exams. 
Participants even stated that writing their reasoning sometimes increased their confidence, as they could clearly see their logical progression. Beyond reinforcing content knowledge, these features also helped students develop metacognitive skills. P13 compared the explanation requirement to their preferred study method of teaching concepts to friends, noting that \textit{``it was almost like I was trying to convince myself why I thought this answer was the best.''} 
P16 reported a notable shift in their study approach beginning with Midterm 2, where they started crafting more elaborate explanations, even for questions they felt uncertain about. They began associating the depth of their explanations with a clearer understanding of concepts—demonstrating increased metacognitive awareness and internal motivation. This change was also reinforced by a noticeable improvement in the quality of AI feedback P16 received: when their explanations were more detailed, the AI responses became more substantive and helpful. 

Despite the benefits, the reflection components were not universally appreciated. 
Some students viewed them as a waste of their time and admitted to writing filler responses, especially when they had absolutely no idea what the answer was. P2 acknowledged the psychological benefits of reflection but found it unnecessary for questions where they were “100\% certain.” P17, who reported having ADHD, found that the requirement was mentally exhausting, particularly when applied to every question. They suggested that it should be optional for straightforward problems, allowing students to focus their energy on more complex ones. The confidence feature also had its detractors. Some students struggled with rating their confidence accurately, often defaulting to a neutral rating on the scale. They rarely selected the highest confidence level, noting that they generally did not feel fully confident in their answers until they received confirmation. P9, who described themselves as an overthinker, found the confidence slider challenging because they could always identify both strengths and weaknesses in their reasoning, leading them to repeatedly select “somewhat confident.”  P19 appreciated that the confidence slider forced them to \textit{``be honest with myself''} rather than instinctively selecting an answer, but they also noted that it made the practice exam take significantly longer to complete. In contrast, P7 found an unexpected benefit in the system—when they marked \textit{``just purely guessing''} in the confidence slider, they received more detailed feedback, which ultimately improved their confidence when retrying those questions. P16, who initially felt stressed about articulating their reasoning, adapted by writing about what confused them, which sometimes led to realizations that they knew more than they had initially thought.

\paragraph{\textbf{Student engagement with feedback and learning resources during review mode:}} 

Students' interaction with AI-generated feedback and their study habits were heavily shaped by the confidence ratings they assigned during their initial attempts. Many students reported that their confidence level determined how thoroughly they engaged with the feedback and whether they consulted additional resources. Student engagement with feedback varied significantly based on the intersection of confidence and correctness. As reported in our survey, students were most likely to thoroughly engage with feedback when their confidence level mismatched the outcome—specifically when high-confidence answers were incorrect or low-confidence answers were correct. P8 noted that when they got a low-confidence question correct, they reviewed the feedback to ensure that their reasoning aligned with the correct answer, rather than assuming they had simply guessed correctly. In contrast, if they were highly confident and answered correctly, they often skimmed the feedback for confirmation or disregarded it entirely. Finally, moments where students felt completely certain about an answer but discovered they were wrong prompted the strongest engagement—these instances led them to scrutinize the feedback, revisit textbook readings, and seek further clarification from external sources to resolve their misconceptions. However, not all students found the confidence ratings beneficial during review, particularly those who struggled with accurately assessing their own confidence from the outset. A smaller subset of students largely disregarded their confidence ratings altogether, prioritizing whether their answers were correct or incorrect instead. Notably, these were often the same students who had found the confidence rating requirement cumbersome during the initial attempt phase.

Students had mixed perceptions of the AI-generated feedback, with some finding it highly useful for correcting misconceptions and others criticizing its limitations. P1 explained that the AI feedback was most helpful when they already had some understanding of the concept because it provided additional details that reinforced their knowledge. For some students, the AI feedback served as a scaffold to refine their reasoning. P5 appreciated that the feedback did not simply confirm whether an answer was right or wrong but also analyzed their explanation, pointing out gaps in their reasoning. They found this especially valuable when the feedback suggested alternative ways of thinking about a problem. P10 recalled an instance where they received AI feedback that critiqued their explanation despite selecting the correct answer, which helped them realize that their reasoning was flawed and could lead to mistakes in future assessments. P12 found it particularly helpful in making interdisciplinary connections, as they were simultaneously taking a chemistry course and appreciated how the AI linked chemistry concepts to biological applications. Some students described how the AI’s structured explanations helped them \textit{``get to the answer faster rather than thinking about it in an abstract way.''} However, others expressed frustration when the feedback lacked depth. They argued that a direct three-sentence summary explaining why an answer was correct or incorrect would have been most useful. P7 similarly noted that while they valued guiding (socratic-style) questions in feedback, they sometimes found them too vague, leading them to search for additional clarification in other resources. Students  reported the AI-generated text was sometimes too dense and difficult to read in the small feedback window, making it less engaging.

Students had varied opinions of the AI's qualities in the feedback. P15 expressed skepticism toward AI-generated feedback, preferring clear disclosure that the responses were AI-generated rather than potentially mistaken for instructor-written explanations. Their distrust stemmed from the \textit{``robotic''} tone, occasional inaccuracies, and an overall uncertainty about the reliability of the feedback. There was a stark contrast in other students' receptivity, with others appreciating the tone and quality of the AI-generated feedback. 
w
One recurring suggestion among students was the inclusion of human-verified explanations. P9 proposed incorporating short instructor- or TA-led videos to accompany the AI feedback, arguing that a concise one- to three-minute explanation would be more digestible than lengthy text-based responses. They also pointed out that students sometimes turn to external sources like YouTube because they prefer human explanations over AI-generated ones.

Students exhibited varying engagement levels with the textbook links accompanying the feedback. Some, like P3, frequently used the links as a reference, while others, like P6, admitted that they rarely clicked on them despite recognizing their potential utility. P1 recommended that textbook links should lead directly to specific subsections rather than entire chapters. 
P7 echoed this, explaining that they often found themselves scrolling through lengthy readings to locate relevant information. Several students reported turning to external resources when the provided textbook links or AI feedback were insufficient. Some students corroborated the AI feedback with at least two sources (one of which was the textbook links), ensuring that they fully understood the concept before moving on.

\paragraph{\textbf{Perceived tool effects on midterm performance knowledge transfer:}} 
Students overwhelmingly found the tool to be instrumental in their preparation for the midterm. However, perspectives varied regarding the extent and nature of their effectiveness. 

Students discussed how they instinctively recalled and reused strategies (they had picked up from the tool and its feedback) when attempting the actual midterm. Multiple students reported encountering identical or highly similar questions in the midterm, which allowed them to leverage their prior mistakes and apply the correct reasoning. For P9, pattern recognition played a significant role in rectifying misconceptions. They remembered a question they had answered incorrectly twice on the practice exam and used the feedback to solidify the necessary procedural steps. When encountering a similar question on the midterm, they systematically applied the same analytical approach, demonstrating a transfer of conceptual understanding rather than mere memorization. P12 shared that they often struggled with confusion around really complex concepts. The AI feedback played a crucial role in prompting them to double-check their understanding, ensuring they reviewed the right information instead of reinforcing misconceptions. This extra review not only clarified tricky details but gave them greater confidence when answering similar questions on the actual midterm. 

Finally, students reported various study and test-taking strategies that emerged from using the tool. P16 adopted a structured approach by answering the questions they were most confident in first before revisiting the more uncertain ones, attributing this strategy to the tool’s emphasis on confidence reflection. P15, who had extended exam time accommodations due to stress-related challenges, found that practicing with the tool helped with pace: 

\begin{quote}
\textit{``I start stressing out if I notice that the exam time is really short...But I noticed myself slowing down because of the tool...I have a hard time slowing down on exams because of the time limit, so having the [tool] practice exam beforehand for thinking through my answers, really helped...It helps you learn, like, `Okay, this is a bad habit of mine. Let's slow it down and unlearn that.'''}
\end{quote}

P19 adopted the explanation strategies from the practice tool by deliberately writing out their thought process for each question on the midterm. However, they worried that if they initially rationalized an incorrect answer during practice, it could inadvertently reinforce a misconception and lead them to recall the wrong response later. Despite these nuances, students overwhelmingly agreed that engaging with the practice tool significantly contributed to their increased confidence for the midterm. 

These behaviors carried over into future courses, even after students no longer had access to our tool. P16 spoke about continuing the practice of gauging their confidence on low-certainty questions and writing out their reasoning in detail as a way to reflect on their learning. They described this as a valuable method for tracking their progress in their new courses.
Similarly, P4 shared that they began using ChatGPT in their new classes to replicate the feedback loop provided by our tool. They tackled new practice problems by writing out full justifications for their answers, and then submitting those to GPT for critique. This allowed them to reframe complex concepts in their own words and receive targeted feedback on the precision and completeness of their understanding. 
These examples illustrate not only the cross-contextual transfer and persistence of reflective learning strategies, but also the adaptive repurposing of available AI tools to meet similar cognitive and reflective goals.

\section{Discussion}

\subsection{RQ1: Impact on Performance}
Our findings present interesting contrasts with prior work in educational feedback and student engagement. While Kulhavy et al.~\cite{kulhavy1989feedback} argued that simple correctness feedback was insufficient, our results showed a more nuanced picture. Across all midterms, we found no statistically significant differences between the various feedback conditions, suggesting that in the presence of structured metacognitive activities, the specific type of feedback may be less important than previously thought. The most robust finding was the strong relationship between student confidence ratings and performance, highlighting the importance of metacognitive elements in the process. This suggests that requiring students to engage in structured self-explanation and confidence assessment may itself create sufficient cognitive engagement to enhance learning, potentially diminishing the relative impact of feedback types.

Furthermore, our analysis of students who initially scored in the bottom 20\% on their practice exams revealed a particularly interesting trend: while the overall impact of combined AI and textbook feedback (Condition 4) was modest, these struggling students showed marginally significant improvements with this condition ($\beta = 0.049$, $p = 0.067$). This suggests that more comprehensive feedback approaches may be especially beneficial for supporting students needing additional assistance, showing the potential value of differentiated feedback based on student performance levels.

% \balance

\subsection{RQ2: Impact on Engagement and Learning Behaviors}
Our findings reinforce existing literature on students' motivation in academic reading and assessment engagement. Prior research has shown that students often prioritize exam performance over deep engagement with course text materials \cite{sappington2002two, howard2004just}. This study further highlights that students' primary motivation for using the practice tool was to improve their midterm performance, aligning with previous findings that students are driven more by assessment outcomes than intrinsic engagement with course content \cite{vafeas2013attitudes, french2015textbook}.

Traditional practice exams can only tell students whether their answers are correct or not, which means students are only incentivized to provide answers.   But an AI feedback tool that adapts its responses to students' explanations and confidence can incentivize students to invest in structured reflection about explanations and confidence, creating an assessment-driven learning experience that encourages deeper cognitive engagement. The positive reception is evidenced by the 76\% voluntary adoption rate for midterm 2, showing students found value in the tool beyond course requirements.

% SELF REGULATION BEHAVIORS SUMMARY + LITERATURE CONNECTION
The structured self-explanation and confidence rating requirements in the tool reflect established pedagogical approaches that promote metacognitive awareness \cite{derryberry2008relationships, tomasek2009critical}. Many students reported that articulating their reasoning forced them to reconsider assumptions, distinguish between educated guesses and true understanding, and refine their problem-solving strategies. This supports previous research demonstrating that self-explanation and question-generation enhance comprehension and retention \cite{smith2010students, van2006study, marbach2000can}. However, our findings also reveal potential drawbacks, particularly for students who found the requirement mentally exhausting or redundant when their confidence was on the extreme ends of the spectrum (\textit{``very confident''} or \textit{``basically guessing''}). Thus, for students with minimal confidence, providing an option to bypass explanation requirements could be beneficial. Many students instinctively turned to textbook links before engaging with AI feedback to build foundational knowledge in such cases, aligning with existing research on confidence and feedback receptivity \cite{hattie2007power, kulhavy1989feedback}.

% TEXTBOOK NAVIGATION AND AI FEEDBACK SCAFFOLDS + FUTURE DIRECTIONS 
The AI-generated feedback adapted to students’ confidence levels and explanations and played a critical role in steering students towards the textbook. Students reported verifying AI feedback against human-authored content, reflecting AI skepticism \cite{tick2024exploring}, but also enabling clarification of misconceptions and supplementing less directive AI feedback. This suggests future systems should support layered scaffolding that balances open-ended prompting with direct instructional resources, and facilitates verification behaviors to promote critical AI literacy. 
The observed transfer of reasoning and pacing strategies from the practice tool to the midterm underscores the potential of structured AI interactions to influence students' metacognitive behaviors. This aligns with prior work on value of practice testing as a learning strategy \cite{carpenter2012testing, roediger2011critical}.  Some students, however, noted concerns that articulating reasoning early in the process might sometimes lead to reinforcement of incorrect thinking. Future systems could address this by including iterative feedback loops that prompt students to revisit and refine their reasoning over time. Designing AI feedback supporting both immediate understanding and sustained, reflective learning would help.

% TEXTBOOK GRANULARITY + FUTURE DIRECTIONS
While some students found the integration of AI and textbook references helpful for mapping instructional materials to exam questions \cite{biggs2011train, chipperfield2022embedding}, others expressed frustration when referred to dense textbook sections. This aligns with prior findings that students often prefer digestible explanations over extensive references \cite{howard2004just, brost2006student}. Future iterations could enhance usability through more targeted textbook linking, AI-generated summaries of relevant sections, or incorporating human-verified content like short instructor videos. The student explanations collected can inform refined AI prompts, enabling more personalized feedback aligned with student reasoning patterns. Students found answering identical questions on multiple attempts ineffective; varying question formats on reattempts would better support knowledge transfer and retention \cite{roediger2011critical}. 

% RETHINKING FUTURE RE-DESIGN OF TOOLS
More fundamentally, our findings point to a possible redesign opportunity for educational platforms. Rather than positioning metacognitive components as supporting features for content delivery, future systems might invert this relationship—making metacognitive development the explicit design goal while AI feedback serves as scaffolding. Such systems could analyze confidence patterns over time, provide targeted feedback on explanation quality, and gradually reduce structured support as students develop independent reflection. This would leverage technology to cultivate transferable critical thinking skills that persist beyond specific course contexts.

\section{Limitations}
Our study focused on a single undergraduate biology course, potentially limiting generalizability across STEM disciplines. While GPT-4o proved effective for feedback generation, other AI approaches might offer different benefits. Self-reported survey data has inherent biases, and the lack of baseline biology knowledge assessment made it difficult to control for varying levels of prior subject expertise.

\section{Conclusion}
Our work demonstrates how AI-enhanced practice systems can support learning at scale through careful integration of technology and pedagogical principles. While textbook references alone showed modest performance gains, the system's greater impact emerged in transforming students' learning behaviors and metacognitive strategies. The high textbook engagement rate and successful adoption of self-assessment practices suggest that contextual, just-in-time support can effectively motivate student engagement with course materials. As institutions explore AI integration in education, our results emphasize designing systems that enhance rather than replace traditional learning resources, while supporting the development of sustainable study practices.

\section{Acknowledgments}
We thank Mason Johnstone for contributing practice exam question drafts from the course for consideration.

% \begin{acks}
% \end{acks}

\bibliographystyle{ACM-Reference-Format}
\balance
\bibliography{sample-base}


\begin{thebibliography}{57}

%%% the Web address.

\ifx \showCODEN    \undefined \def \showCODEN     #1{\unskip}     \fi
\ifx \showISBNx    \undefined \def \showISBNx     #1{\unskip}     \fi
\ifx \showISBNxiii \undefined \def \showISBNxiii  #1{\unskip}     \fi
\ifx \showISSN     \undefined \def \showISSN      #1{\unskip}     \fi
\ifx \showLCCN     \undefined \def \showLCCN      #1{\unskip}     \fi
\ifx \shownote     \undefined \def \shownote      #1{#1}          \fi
\ifx \showarticletitle \undefined \def \showarticletitle #1{#1}   \fi
\ifx \showURL      \undefined \def \showURL       {\relax}        \fi
% invisible to TeX
\providecommand\bibfield[2]{#2}
\providecommand\bibinfo[2]{#2}
\providecommand\natexlab[1]{#1}
\providecommand\showeprint[2][]{arXiv:#2}

\bibitem[Ahmad and Ma(2024)]%
        {ahmad1}
\bibfield{author}{\bibinfo{person}{Mak Ahmad} {and} \bibinfo{person}{Kwan-Liu Ma}.} \bibinfo{year}{2024}\natexlab{}.
\newblock \showarticletitle{{More Than Chatting: Conversational LLMs for Enhancing Data Visualization Competencies}}. In \bibinfo{booktitle}{\emph{EuroVis 2024 - Education Papers}}, \bibfield{editor}{\bibinfo{person}{Elif~E. Firat}, \bibinfo{person}{Robert~S. Laramee}, {and} \bibinfo{person}{Nicklas~Sindelv Andersen}} (Eds.). \bibinfo{publisher}{The Eurographics Association}.
\newblock
\showISBNx{978-3-03868-257-8}
\href{https://doi.org/10.2312/eved.20241056}{doi:\nolinkurl{10.2312/eved.20241056}}


\bibitem[Biggs and Tang(2011)]%
        {biggs2011train}
\bibfield{author}{\bibinfo{person}{John Biggs} {and} \bibinfo{person}{Catherine Tang}.} \bibinfo{year}{2011}\natexlab{}.
\newblock \showarticletitle{Train-the-trainers: Implementing outcomes-based teaching and learning in Malaysian higher education.}
\newblock \bibinfo{journal}{\emph{Malaysian Journal of Learning and Instruction}}  \bibinfo{volume}{8} (\bibinfo{year}{2011}), \bibinfo{pages}{1--19}.
\newblock


\bibitem[Brost and Bradley(2006)]%
        {brost2006student}
\bibfield{author}{\bibinfo{person}{Brian Brost} {and} \bibinfo{person}{Karen Bradley}.} \bibinfo{year}{2006}\natexlab{}.
\newblock \showarticletitle{Student compliance with assigned reading: A case study}.
\newblock \bibinfo{journal}{\emph{Journal of the Scholarship of Teaching and Learning}} (\bibinfo{year}{2006}), \bibinfo{pages}{101--111}.
\newblock


\bibitem[Butler et~al\mbox{.}(2007)]%
        {butler2007effect}
\bibfield{author}{\bibinfo{person}{Andrew~C Butler}, \bibinfo{person}{Jeffrey~D Karpicke}, {and} \bibinfo{person}{Henry~L Roediger~III}.} \bibinfo{year}{2007}\natexlab{}.
\newblock \showarticletitle{The effect of type and timing of feedback on learning from multiple-choice tests.}
\newblock \bibinfo{journal}{\emph{Journal of Experimental Psychology: Applied}} \bibinfo{volume}{13}, \bibinfo{number}{4} (\bibinfo{year}{2007}), \bibinfo{pages}{273}.
\newblock


\bibitem[Butler and Winne(1995)]%
        {butler1995feedback}
\bibfield{author}{\bibinfo{person}{Deborah~L Butler} {and} \bibinfo{person}{Philip~H Winne}.} \bibinfo{year}{1995}\natexlab{}.
\newblock \showarticletitle{Feedback and self-regulated learning: A theoretical synthesis}.
\newblock \bibinfo{journal}{\emph{Review of educational research}} \bibinfo{volume}{65}, \bibinfo{number}{3} (\bibinfo{year}{1995}), \bibinfo{pages}{245--281}.
\newblock


\bibitem[Carpenter(2012)]%
        {carpenter2012testing}
\bibfield{author}{\bibinfo{person}{Shana~K Carpenter}.} \bibinfo{year}{2012}\natexlab{}.
\newblock \showarticletitle{Testing enhances the transfer of learning}.
\newblock \bibinfo{journal}{\emph{Current directions in psychological science}} \bibinfo{volume}{21}, \bibinfo{number}{5} (\bibinfo{year}{2012}), \bibinfo{pages}{279--283}.
\newblock


\bibitem[Chang(2010)]%
        {chang2010effects}
\bibfield{author}{\bibinfo{person}{Mei-Mei Chang}.} \bibinfo{year}{2010}\natexlab{}.
\newblock \showarticletitle{Effects of self-monitoring on web-based language learner's performance and motivation}.
\newblock \bibinfo{journal}{\emph{Calico Journal}} \bibinfo{volume}{27}, \bibinfo{number}{2} (\bibinfo{year}{2010}), \bibinfo{pages}{298--310}.
\newblock


\bibitem[Charmaz(2008)]%
        {charmaz2008grounded}
\bibfield{author}{\bibinfo{person}{Kathy Charmaz}.} \bibinfo{year}{2008}\natexlab{}.
\newblock \showarticletitle{Grounded theory as an emergent method}.
\newblock \bibinfo{journal}{\emph{Handbook of emergent methods}}  \bibinfo{volume}{155} (\bibinfo{year}{2008}), \bibinfo{pages}{172}.
\newblock


\bibitem[Chipperfield et~al\mbox{.}(2022)]%
        {chipperfield2022embedding}
\bibfield{author}{\bibinfo{person}{Grace Chipperfield}, \bibinfo{person}{Lauren Butterworth}, {and} \bibinfo{person}{Pablo Munguia}.} \bibinfo{year}{2022}\natexlab{}.
\newblock \showarticletitle{Embedding resources into digital assessment rubrics: Bringing academic support directly to students}.
\newblock \bibinfo{journal}{\emph{Journal of Academic Language and Learning}} \bibinfo{volume}{16}, \bibinfo{number}{1} (\bibinfo{year}{2022}), \bibinfo{pages}{C1--C11}.
\newblock


\bibitem[Clump et~al\mbox{.}(2004)]%
        {clump2004extent}
\bibfield{author}{\bibinfo{person}{Michael~A Clump}, \bibinfo{person}{Heather Bauer}, {and} \bibinfo{person}{Catherine Bradley}.} \bibinfo{year}{2004}\natexlab{}.
\newblock \showarticletitle{The extent to which psychology students read textbooks: a multiple class analysis of reading across the psychology curriculum.}
\newblock \bibinfo{journal}{\emph{Journal of Instructional Psychology}} \bibinfo{volume}{31}, \bibinfo{number}{3} (\bibinfo{year}{2004}), \bibinfo{pages}{227--233}.
\newblock


\bibitem[Connor-Greene(2000)]%
        {connor2000assessing}
\bibfield{author}{\bibinfo{person}{Patricia~A Connor-Greene}.} \bibinfo{year}{2000}\natexlab{}.
\newblock \showarticletitle{Assessing and promoting student learning: Blurring the line between teaching and testing}.
\newblock \bibinfo{journal}{\emph{Teaching of Psychology}} \bibinfo{volume}{27}, \bibinfo{number}{2} (\bibinfo{year}{2000}), \bibinfo{pages}{84--88}.
\newblock


\bibitem[Corbin et~al\mbox{.}(1990)]%
        {corbin1990basics}
\bibfield{author}{\bibinfo{person}{Juliet Corbin} {et~al\mbox{.}}} \bibinfo{year}{1990}\natexlab{}.
\newblock \showarticletitle{Basics of qualitative research grounded theory procedures and techniques}.
\newblock  (\bibinfo{year}{1990}).
\newblock


\bibitem[Dai et~al\mbox{.}(2024)]%
        {dai2024assessing}
\bibfield{author}{\bibinfo{person}{Wei Dai}, \bibinfo{person}{Yi-Shan Tsai}, \bibinfo{person}{Jionghao Lin}, \bibinfo{person}{Ahmad Aldino}, \bibinfo{person}{Hua Jin}, \bibinfo{person}{Tongguang Li}, \bibinfo{person}{Dragan Ga{\v{s}}evi{\'c}}, {and} \bibinfo{person}{Guanliang Chen}.} \bibinfo{year}{2024}\natexlab{}.
\newblock \showarticletitle{Assessing the proficiency of large language models in automatic feedback generation: An evaluation study}.
\newblock \bibinfo{journal}{\emph{Computers and Education: Artificial Intelligence}}  \bibinfo{volume}{7} (\bibinfo{year}{2024}), \bibinfo{pages}{100299}.
\newblock


\bibitem[Derryberry and Wininger(2008)]%
        {derryberry2008relationships}
\bibfield{author}{\bibinfo{person}{W~Pitt Derryberry} {and} \bibinfo{person}{Steven~R Wininger}.} \bibinfo{year}{2008}\natexlab{}.
\newblock \showarticletitle{Relationships among Textbook Usage and Cognitive-Motivational Constructs.}
\newblock \bibinfo{journal}{\emph{Teaching Educational Psychology}} \bibinfo{volume}{3}, \bibinfo{number}{2} (\bibinfo{year}{2008}), \bibinfo{pages}{n2}.
\newblock


\bibitem[Durwin and Sherman(2008)]%
        {durwin2008does}
\bibfield{author}{\bibinfo{person}{Cheryl~Cisero Durwin} {and} \bibinfo{person}{William~M Sherman}.} \bibinfo{year}{2008}\natexlab{}.
\newblock \showarticletitle{Does choice of college textbook make a difference in students' comprehension?}
\newblock \bibinfo{journal}{\emph{College teaching}} \bibinfo{volume}{56}, \bibinfo{number}{1} (\bibinfo{year}{2008}), \bibinfo{pages}{28--34}.
\newblock


\bibitem[French et~al\mbox{.}(2015)]%
        {french2015textbook}
\bibfield{author}{\bibinfo{person}{Michelle French}, \bibinfo{person}{Franco Taverna}, \bibinfo{person}{Melody Neumann}, \bibinfo{person}{Lena Paulo~Kushnir}, \bibinfo{person}{Jason Harlow}, \bibinfo{person}{David Harrison}, {and} \bibinfo{person}{Ruxandra Serbanescu}.} \bibinfo{year}{2015}\natexlab{}.
\newblock \showarticletitle{Textbook use in the sciences and its relation to course performance}.
\newblock \bibinfo{journal}{\emph{College Teaching}} \bibinfo{volume}{63}, \bibinfo{number}{4} (\bibinfo{year}{2015}), \bibinfo{pages}{171--177}.
\newblock


\bibitem[Geerling et~al\mbox{.}(2023)]%
        {geerling2023chatgpt}
\bibfield{author}{\bibinfo{person}{Wayne Geerling}, \bibinfo{person}{G~Dirk Mateer}, \bibinfo{person}{Jadrian Wooten}, {and} \bibinfo{person}{Nikhil Damodaran}.} \bibinfo{year}{2023}\natexlab{}.
\newblock \showarticletitle{Is ChatGPT smarter than a student in principles of economics}.
\newblock \bibinfo{journal}{\emph{Available at SSRN}}  \bibinfo{volume}{4356034} (\bibinfo{year}{2023}).
\newblock


\bibitem[Hatteberg and Steffy(2013)]%
        {hatteberg2013increasing}
\bibfield{author}{\bibinfo{person}{Sarah~J Hatteberg} {and} \bibinfo{person}{Kody Steffy}.} \bibinfo{year}{2013}\natexlab{}.
\newblock \showarticletitle{Increasing reading compliance of undergraduates: An evaluation of compliance methods}.
\newblock \bibinfo{journal}{\emph{Teaching Sociology}} \bibinfo{volume}{41}, \bibinfo{number}{4} (\bibinfo{year}{2013}), \bibinfo{pages}{346--352}.
\newblock


\bibitem[Hattie and Timperley(2007)]%
        {hattie2007power}
\bibfield{author}{\bibinfo{person}{John Hattie} {and} \bibinfo{person}{Helen Timperley}.} \bibinfo{year}{2007}\natexlab{}.
\newblock \showarticletitle{The power of feedback}.
\newblock \bibinfo{journal}{\emph{Review of educational research}} \bibinfo{volume}{77}, \bibinfo{number}{1} (\bibinfo{year}{2007}), \bibinfo{pages}{81--112}.
\newblock


\bibitem[Heiner et~al\mbox{.}(2014)]%
        {heiner2014preparing}
\bibfield{author}{\bibinfo{person}{Cynthia~E Heiner}, \bibinfo{person}{Amanda~I Banet}, {and} \bibinfo{person}{Carl Wieman}.} \bibinfo{year}{2014}\natexlab{}.
\newblock \showarticletitle{Preparing students for class: How to get 80\% of students reading the textbook before class}.
\newblock \bibinfo{journal}{\emph{American Journal of Physics}} \bibinfo{volume}{82}, \bibinfo{number}{10} (\bibinfo{year}{2014}), \bibinfo{pages}{989--996}.
\newblock


\bibitem[Hoeft(2012)]%
        {hoeft2012university}
\bibfield{author}{\bibinfo{person}{Mary~E Hoeft}.} \bibinfo{year}{2012}\natexlab{}.
\newblock \showarticletitle{Why university students don't read: What professors can do to increase compliance}.
\newblock \bibinfo{journal}{\emph{International journal for the scholarship of teaching and learning}} \bibinfo{volume}{6}, \bibinfo{number}{2} (\bibinfo{year}{2012}), \bibinfo{pages}{12}.
\newblock


\bibitem[Howard(2004)]%
        {howard2004just}
\bibfield{author}{\bibinfo{person}{Jay~R Howard}.} \bibinfo{year}{2004}\natexlab{}.
\newblock \showarticletitle{Just-in-time teaching in sociology or how I convinced my students to actually read the assignment}.
\newblock \bibinfo{journal}{\emph{Teaching Sociology}} \bibinfo{volume}{32}, \bibinfo{number}{4} (\bibinfo{year}{2004}), \bibinfo{pages}{385--390}.
\newblock


\bibitem[Howard et~al\mbox{.}(2018)]%
        {howard2018academic}
\bibfield{author}{\bibinfo{person}{Pamela~J Howard}, \bibinfo{person}{Meg Gorzycki}, \bibinfo{person}{Geoffrey Desa}, {and} \bibinfo{person}{Diane~D Allen}.} \bibinfo{year}{2018}\natexlab{}.
\newblock \showarticletitle{Academic reading: Comparing students’ and faculty perceptions of its value, practice, and pedagogy}.
\newblock \bibinfo{journal}{\emph{Journal of College Reading and Learning}} \bibinfo{volume}{48}, \bibinfo{number}{3} (\bibinfo{year}{2018}), \bibinfo{pages}{189--209}.
\newblock


\bibitem[Jensen et~al\mbox{.}(2024)]%
        {jensen2024generative}
\bibfield{author}{\bibinfo{person}{Lasse~X Jensen}, \bibinfo{person}{Alexandra Buhl}, \bibinfo{person}{Anjali Sharma}, {and} \bibinfo{person}{Margaret Bearman}.} \bibinfo{year}{2024}\natexlab{}.
\newblock \showarticletitle{Generative AI and higher education: a review of claims from the first months of ChatGPT}.
\newblock \bibinfo{journal}{\emph{Higher Education}} (\bibinfo{year}{2024}), \bibinfo{pages}{1--17}.
\newblock


\bibitem[Johnson and Kiviniemi(2009)]%
        {johnson2009effect}
\bibfield{author}{\bibinfo{person}{Bethany~C Johnson} {and} \bibinfo{person}{Marc~T Kiviniemi}.} \bibinfo{year}{2009}\natexlab{}.
\newblock \showarticletitle{The effect of online chapter quizzes on exam performance in an undergraduate social psychology course}.
\newblock \bibinfo{journal}{\emph{Teaching of Psychology}} \bibinfo{volume}{36}, \bibinfo{number}{1} (\bibinfo{year}{2009}), \bibinfo{pages}{33--37}.
\newblock


\bibitem[Klymkowsky and Cooper(2024)]%
        {klymkowsky2024end}
\bibfield{author}{\bibinfo{person}{Michael Klymkowsky} {and} \bibinfo{person}{Melanie~M Cooper}.} \bibinfo{year}{2024}\natexlab{}.
\newblock \showarticletitle{The end of multiple choice tests: using AI to enhance assessment}.
\newblock \bibinfo{journal}{\emph{arXiv preprint arXiv:2406.07481}} (\bibinfo{year}{2024}).
\newblock


\bibitem[Kulhavy and Stock(1989)]%
        {kulhavy1989feedback}
\bibfield{author}{\bibinfo{person}{Raymond~W Kulhavy} {and} \bibinfo{person}{William~A Stock}.} \bibinfo{year}{1989}\natexlab{}.
\newblock \showarticletitle{Feedback in written instruction: The place of response certitude}.
\newblock \bibinfo{journal}{\emph{Educational psychology review}}  \bibinfo{volume}{1} (\bibinfo{year}{1989}), \bibinfo{pages}{279--308}.
\newblock


\bibitem[Kulhavy et~al\mbox{.}(1985)]%
        {kulhavy1985feedback}
\bibfield{author}{\bibinfo{person}{Raymond~W Kulhavy}, \bibinfo{person}{Mary~T White}, \bibinfo{person}{Bruce~W Topp}, \bibinfo{person}{Ann~L Chan}, {and} \bibinfo{person}{James Adams}.} \bibinfo{year}{1985}\natexlab{}.
\newblock \showarticletitle{Feedback complexity and corrective efficiency}.
\newblock \bibinfo{journal}{\emph{Contemporary educational psychology}} \bibinfo{volume}{10}, \bibinfo{number}{3} (\bibinfo{year}{1985}), \bibinfo{pages}{285--291}.
\newblock


\bibitem[Kulik and Kulik(1988)]%
        {kulik1988timing}
\bibfield{author}{\bibinfo{person}{James~A Kulik} {and} \bibinfo{person}{Chen-Lin~C Kulik}.} \bibinfo{year}{1988}\natexlab{}.
\newblock \showarticletitle{Timing of feedback and verbal learning}.
\newblock \bibinfo{journal}{\emph{Review of educational research}} \bibinfo{volume}{58}, \bibinfo{number}{1} (\bibinfo{year}{1988}), \bibinfo{pages}{79--97}.
\newblock


\bibitem[Kumar et~al\mbox{.}(2024)]%
        {kumar2024supporting}
\bibfield{author}{\bibinfo{person}{Harsh Kumar}, \bibinfo{person}{Ruiwei Xiao}, \bibinfo{person}{Benjamin Lawson}, \bibinfo{person}{Ilya Musabirov}, \bibinfo{person}{Jiakai Shi}, \bibinfo{person}{Xinyuan Wang}, \bibinfo{person}{Huayin Luo}, \bibinfo{person}{Joseph~Jay Williams}, \bibinfo{person}{Anna~N Rafferty}, \bibinfo{person}{John Stamper}, {et~al\mbox{.}}} \bibinfo{year}{2024}\natexlab{}.
\newblock \showarticletitle{Supporting self-reflection at scale with large language models: Insights from randomized field experiments in classrooms}. In \bibinfo{booktitle}{\emph{Proceedings of the eleventh ACM conference on learning@ scale}}. \bibinfo{pages}{86--97}.
\newblock


\bibitem[Lehmann et~al\mbox{.}(2024)]%
        {lehmann2024ai}
\bibfield{author}{\bibinfo{person}{Matthias Lehmann}, \bibinfo{person}{Philipp~B Cornelius}, {and} \bibinfo{person}{Fabian~J Sting}.} \bibinfo{year}{2024}\natexlab{}.
\newblock \showarticletitle{AI Meets the Classroom: When Does ChatGPT Harm Learning?}
\newblock \bibinfo{journal}{\emph{arXiv preprint arXiv:2409.09047}} (\bibinfo{year}{2024}).
\newblock


\bibitem[Leite and Blanco(2020)]%
        {leite2020effects}
\bibfield{author}{\bibinfo{person}{Abe Leite} {and} \bibinfo{person}{Sa{\'u}l~A Blanco}.} \bibinfo{year}{2020}\natexlab{}.
\newblock \showarticletitle{Effects of human vs. automatic feedback on students' understanding of AI concepts and programming style}. In \bibinfo{booktitle}{\emph{Proceedings of the 51st ACM Technical Symposium on Computer Science Education}}. \bibinfo{pages}{44--50}.
\newblock


\bibitem[Lockhart and Soliday(2016)]%
        {lockhart2016critical}
\bibfield{author}{\bibinfo{person}{Tara Lockhart} {and} \bibinfo{person}{Mary Soliday}.} \bibinfo{year}{2016}\natexlab{}.
\newblock \showarticletitle{The critical place of reading in writing transfer (and beyond): A report of student experiences}.
\newblock \bibinfo{journal}{\emph{Pedagogy}} \bibinfo{volume}{16}, \bibinfo{number}{1} (\bibinfo{year}{2016}), \bibinfo{pages}{23--37}.
\newblock


\bibitem[Lysakowski and Walberg(1981)]%
        {lysakowski1981classroom}
\bibfield{author}{\bibinfo{person}{Richard~S Lysakowski} {and} \bibinfo{person}{Herbert~J Walberg}.} \bibinfo{year}{1981}\natexlab{}.
\newblock \showarticletitle{Classroom reinforcement and learning: A quantitative synthesis}.
\newblock \bibinfo{journal}{\emph{The Journal of Educational Research}} \bibinfo{volume}{75}, \bibinfo{number}{2} (\bibinfo{year}{1981}), \bibinfo{pages}{69--77}.
\newblock


\bibitem[Maher and Mitchell(2010)]%
        {maher2010m}
\bibfield{author}{\bibinfo{person}{JaneMaree Maher} {and} \bibinfo{person}{Jennifer Mitchell}.} \bibinfo{year}{2010}\natexlab{}.
\newblock \showarticletitle{I'm not sure what to do! Learning experiences in the humanities and social sciences}.
\newblock \bibinfo{journal}{\emph{Issues in Educational Research}} \bibinfo{volume}{20}, \bibinfo{number}{2} (\bibinfo{year}{2010}), \bibinfo{pages}{137}.
\newblock


\bibitem[Marbach-Ad and Sokolove(2000)]%
        {marbach2000can}
\bibfield{author}{\bibinfo{person}{Gili Marbach-Ad} {and} \bibinfo{person}{Phillip~G Sokolove}.} \bibinfo{year}{2000}\natexlab{}.
\newblock \showarticletitle{Can undergraduate biology students learn to ask higher level questions?}
\newblock \bibinfo{journal}{\emph{Journal of Research in Science Teaching: The Official Journal of the National Association for Research in Science Teaching}} \bibinfo{volume}{37}, \bibinfo{number}{8} (\bibinfo{year}{2000}), \bibinfo{pages}{854--870}.
\newblock


\bibitem[Murden and Gillespie(1997)]%
        {murden1997role}
\bibfield{author}{\bibinfo{person}{Teresa Murden} {and} \bibinfo{person}{Cindy~S Gillespie}.} \bibinfo{year}{1997}\natexlab{}.
\newblock \showarticletitle{The role of textbooks and reading in content area classrooms: What are teachers and students saying}.
\newblock \bibinfo{journal}{\emph{Exploring literacy}} (\bibinfo{year}{1997}), \bibinfo{pages}{87--96}.
\newblock


\bibitem[Nikolic et~al\mbox{.}(2023)]%
        {nikolic2023chatgpt}
\bibfield{author}{\bibinfo{person}{Sasha Nikolic}, \bibinfo{person}{Scott Daniel}, \bibinfo{person}{Rezwanul Haque}, \bibinfo{person}{Marina Belkina}, \bibinfo{person}{Ghulam~M Hassan}, \bibinfo{person}{Sarah Grundy}, \bibinfo{person}{Sarah Lyden}, \bibinfo{person}{Peter Neal}, {and} \bibinfo{person}{Caz Sandison}.} \bibinfo{year}{2023}\natexlab{}.
\newblock \showarticletitle{ChatGPT versus engineering education assessment: a multidisciplinary and multi-institutional benchmarking and analysis of this generative artificial intelligence tool to investigate assessment integrity}.
\newblock \bibinfo{journal}{\emph{European Journal of Engineering Education}} \bibinfo{volume}{48}, \bibinfo{number}{4} (\bibinfo{year}{2023}), \bibinfo{pages}{559--614}.
\newblock


\bibitem[Nolen(1996)]%
        {nolen1996study}
\bibfield{author}{\bibinfo{person}{Susan~Bobbitt Nolen}.} \bibinfo{year}{1996}\natexlab{}.
\newblock \showarticletitle{Why study? How reasons for learning influence strategy selection}.
\newblock \bibinfo{journal}{\emph{Educational Psychology Review}}  \bibinfo{volume}{8} (\bibinfo{year}{1996}), \bibinfo{pages}{335--355}.
\newblock


\bibitem[Roediger and Butler(2011)]%
        {roediger2011critical}
\bibfield{author}{\bibinfo{person}{Henry~L Roediger} {and} \bibinfo{person}{Andrew~C Butler}.} \bibinfo{year}{2011}\natexlab{}.
\newblock \showarticletitle{The critical role of retrieval practice in long-term retention}.
\newblock \bibinfo{journal}{\emph{Trends in cognitive sciences}} \bibinfo{volume}{15}, \bibinfo{number}{1} (\bibinfo{year}{2011}), \bibinfo{pages}{20--27}.
\newblock


\bibitem[Rothkopf(1988)]%
        {rothkopf1988perspectives}
\bibfield{author}{\bibinfo{person}{Ernst~Z Rothkopf}.} \bibinfo{year}{1988}\natexlab{}.
\newblock \showarticletitle{Perspectives on study skills training in a realistic instructional economy}.
\newblock In \bibinfo{booktitle}{\emph{Learning and study strategies}}. \bibinfo{publisher}{Elsevier}, \bibinfo{pages}{275--286}.
\newblock


\bibitem[Sappington et~al\mbox{.}(2002)]%
        {sappington2002two}
\bibfield{author}{\bibinfo{person}{John Sappington}, \bibinfo{person}{Kimberly Kinsey}, {and} \bibinfo{person}{Kirk Munsayac}.} \bibinfo{year}{2002}\natexlab{}.
\newblock \showarticletitle{Two studies of reading compliance among college students}.
\newblock \bibinfo{journal}{\emph{Teaching of psychology}} \bibinfo{volume}{29}, \bibinfo{number}{4} (\bibinfo{year}{2002}), \bibinfo{pages}{272--274}.
\newblock


\bibitem[Sengupta(2002)]%
        {sengupta2002developing}
\bibfield{author}{\bibinfo{person}{Sima Sengupta}.} \bibinfo{year}{2002}\natexlab{}.
\newblock \showarticletitle{Developing academic reading at tertiary level: A longitudinal study tracing conceptual change}.
\newblock \bibinfo{journal}{\emph{The reading matrix}} \bibinfo{volume}{2}, \bibinfo{number}{1} (\bibinfo{year}{2002}).
\newblock


\bibitem[Smith et~al\mbox{.}(2010)]%
        {smith2010students}
\bibfield{author}{\bibinfo{person}{Betty~Lou Smith}, \bibinfo{person}{William~G Holliday}, {and} \bibinfo{person}{Homer~W Austin}.} \bibinfo{year}{2010}\natexlab{}.
\newblock \showarticletitle{Students' comprehension of science textbooks using a question-based reading strategy}.
\newblock \bibinfo{journal}{\emph{Journal of Research in Science Teaching: The Official Journal of the National Association for Research in Science Teaching}} \bibinfo{volume}{47}, \bibinfo{number}{4} (\bibinfo{year}{2010}), \bibinfo{pages}{363--379}.
\newblock


\bibitem[Smolansky et~al\mbox{.}(2023)]%
        {smolansky2023educator}
\bibfield{author}{\bibinfo{person}{Adele Smolansky}, \bibinfo{person}{Andrew Cram}, \bibinfo{person}{Corina Raduescu}, \bibinfo{person}{Sandris Zeivots}, \bibinfo{person}{Elaine Huber}, {and} \bibinfo{person}{Rene~F Kizilcec}.} \bibinfo{year}{2023}\natexlab{}.
\newblock \showarticletitle{Educator and student perspectives on the impact of generative AI on assessments in higher education}. In \bibinfo{booktitle}{\emph{Proceedings of the tenth ACM conference on Learning@ Scale}}. \bibinfo{pages}{378--382}.
\newblock


\bibitem[St~Clair-Thompson et~al\mbox{.}(2018)]%
        {st2018exploring}
\bibfield{author}{\bibinfo{person}{Helen St~Clair-Thompson}, \bibinfo{person}{Alison Graham}, {and} \bibinfo{person}{Sara Marsham}.} \bibinfo{year}{2018}\natexlab{}.
\newblock \showarticletitle{Exploring the reading practices of undergraduate students}.
\newblock \bibinfo{journal}{\emph{Education Inquiry}} \bibinfo{volume}{9}, \bibinfo{number}{3} (\bibinfo{year}{2018}), \bibinfo{pages}{284--298}.
\newblock


\bibitem[Starcher and Proffitt(2011)]%
        {starcher2011encouraging}
\bibfield{author}{\bibinfo{person}{Keith Starcher} {and} \bibinfo{person}{Dennis Proffitt}.} \bibinfo{year}{2011}\natexlab{}.
\newblock \showarticletitle{Encouraging Students to Read: What Professors Are (and Aren't) Doing About It.}
\newblock \bibinfo{journal}{\emph{International Journal of Teaching and Learning in Higher Education}} \bibinfo{volume}{23}, \bibinfo{number}{3} (\bibinfo{year}{2011}), \bibinfo{pages}{396--407}.
\newblock


\bibitem[Stokes-Eley(2007)]%
        {stokes2007using}
\bibfield{author}{\bibinfo{person}{Stephanie Stokes-Eley}.} \bibinfo{year}{2007}\natexlab{}.
\newblock \showarticletitle{Using Kolb's experiential learning cycle in chapter presentations}.
\newblock \bibinfo{journal}{\emph{Communication Teacher}} \bibinfo{volume}{21}, \bibinfo{number}{1} (\bibinfo{year}{2007}), \bibinfo{pages}{26--29}.
\newblock


\bibitem[Stutz et~al\mbox{.}(2023)]%
        {stutz2023ch}
\bibfield{author}{\bibinfo{person}{Petra Stutz}, \bibinfo{person}{Maximilian Elixhauser}, \bibinfo{person}{Judith Grubinger-Preiner}, \bibinfo{person}{Vivienne Linner}, \bibinfo{person}{Eva Reibersdorfer-Adelsberger}, \bibinfo{person}{Christoph Traun}, \bibinfo{person}{Gudrun Wallentin}, \bibinfo{person}{Katharina W{\"o}hs}, {and} \bibinfo{person}{Thomas Zuberb{\"u}hler}.} \bibinfo{year}{2023}\natexlab{}.
\newblock \showarticletitle{Ch (e) atGPT? An anecdotal approach addressing the impact of ChatGPT on teaching and learning GIScience}.
\newblock  (\bibinfo{year}{2023}).
\newblock


\bibitem[Tang and Zhan(2021)]%
        {tang2021does}
\bibfield{author}{\bibinfo{person}{Fang Tang} {and} \bibinfo{person}{Peida Zhan}.} \bibinfo{year}{2021}\natexlab{}.
\newblock \showarticletitle{Does diagnostic feedback promote learning? Evidence from a longitudinal cognitive diagnostic assessment}.
\newblock \bibinfo{journal}{\emph{AERA Open}}  \bibinfo{volume}{7} (\bibinfo{year}{2021}), \bibinfo{pages}{23328584211060804}.
\newblock


\bibitem[Tick(2024)]%
        {tick2024exploring}
\bibfield{author}{\bibinfo{person}{Andrea Tick}.} \bibinfo{year}{2024}\natexlab{}.
\newblock \showarticletitle{Exploring ChatGPT's Potential and Concerns in Higher Education}. In \bibinfo{booktitle}{\emph{2024 IEEE 22nd Jubilee international symposium on intelligent systems and informatics (SISY)}}. IEEE, \bibinfo{pages}{000447--000454}.
\newblock


\bibitem[Tomasek(2009)]%
        {tomasek2009critical}
\bibfield{author}{\bibinfo{person}{Terry Tomasek}.} \bibinfo{year}{2009}\natexlab{}.
\newblock \showarticletitle{Critical reading: Using reading prompts to promote active engagement with text.}
\newblock \bibinfo{journal}{\emph{International journal of teaching and learning in higher education}} \bibinfo{volume}{21}, \bibinfo{number}{1} (\bibinfo{year}{2009}), \bibinfo{pages}{127--132}.
\newblock


\bibitem[Vafeas(2013)]%
        {vafeas2013attitudes}
\bibfield{author}{\bibinfo{person}{Mario Vafeas}.} \bibinfo{year}{2013}\natexlab{}.
\newblock \showarticletitle{Attitudes toward, and use of, textbooks among marketing undergraduates: an exploratory study}.
\newblock \bibinfo{journal}{\emph{Journal of Marketing Education}} \bibinfo{volume}{35}, \bibinfo{number}{3} (\bibinfo{year}{2013}), \bibinfo{pages}{245--258}.
\newblock


\bibitem[Van~Blerkom et~al\mbox{.}(2006)]%
        {van2006study}
\bibfield{author}{\bibinfo{person}{Dianna~L Van~Blerkom}, \bibinfo{person}{Malcolm~L Van~Blerkom}, {and} \bibinfo{person}{Sharon Bertsch}.} \bibinfo{year}{2006}\natexlab{}.
\newblock \showarticletitle{Study strategies and generative learning: What works?}
\newblock \bibinfo{journal}{\emph{Journal of College Reading and Learning}} \bibinfo{volume}{37}, \bibinfo{number}{1} (\bibinfo{year}{2006}), \bibinfo{pages}{7--18}.
\newblock


\bibitem[Ward et~al\mbox{.}(2024)]%
        {ward2024analyzing}
\bibfield{author}{\bibinfo{person}{Ben Ward}, \bibinfo{person}{Deepshikha Bhati}, \bibinfo{person}{Fnu Neha}, {and} \bibinfo{person}{Angela Guercio}.} \bibinfo{year}{2024}\natexlab{}.
\newblock \showarticletitle{Analyzing the Impact of AI Tools on Student Study Habits and Academic Performance}.
\newblock \bibinfo{journal}{\emph{arXiv preprint arXiv:2412.02166}} (\bibinfo{year}{2024}).
\newblock


\bibitem[Wisniewski et~al\mbox{.}(2020)]%
        {wisniewski2020power}
\bibfield{author}{\bibinfo{person}{Benedikt Wisniewski}, \bibinfo{person}{Klaus Zierer}, {and} \bibinfo{person}{John Hattie}.} \bibinfo{year}{2020}\natexlab{}.
\newblock \showarticletitle{The power of feedback revisited: A meta-analysis of educational feedback research}.
\newblock \bibinfo{journal}{\emph{Frontiers in psychology}}  \bibinfo{volume}{10} (\bibinfo{year}{2020}), \bibinfo{pages}{487662}.
\newblock


\bibitem[Xia et~al\mbox{.}(2024)]%
        {xia2024scoping}
\bibfield{author}{\bibinfo{person}{Qi Xia}, \bibinfo{person}{Xiaojing Weng}, \bibinfo{person}{Fan Ouyang}, \bibinfo{person}{Tzung~Jin Lin}, {and} \bibinfo{person}{Thomas~KF Chiu}.} \bibinfo{year}{2024}\natexlab{}.
\newblock \showarticletitle{A scoping review on how generative artificial intelligence transforms assessment in higher education}.
\newblock \bibinfo{journal}{\emph{International Journal of Educational Technology in Higher Education}} \bibinfo{volume}{21}, \bibinfo{number}{1} (\bibinfo{year}{2024}), \bibinfo{pages}{40}.
\newblock


\end{thebibliography}

% \appendix

\end{document}